\documentstyle[aps,prl,epsf]{revtex}

\def\be{\begin{equation}}
\def\ee{\end{equation}}
\def\ba{\begin{eqnarray}}
\def\ea{\end{eqnarray}}

\def\C60{A$_x$C$_{60}$}

\def\seX{(TMTSF)$_2$X}
\def\sX{(TMTTF)$_2$X}

\def\spf{(TMTTF)$_2$PF$_6$}
\def\sasf{(TMTTF)$_2$AsF$_6$}

\def\etal{{\it{et al.}}}

\begin{document}

\twocolumn[\hsize\textwidth\columnwidth\hsize\csname@twocolumnfalse\endcsname 

\title{Charge ordering in the TMTTF family of molecular conductors}

\author{D.~S.~Chow$^1$, F. Zamborszky$^2$, B. Alavi$^1$, D.~J.~Tantillo$^3$, A. Baur$^3$, C.~A.~Merlic$^3$, S.~E.~Brown$^1$}
\address{
1) Department of Physics and Astronomy, 
UCLA,
Los Angeles, CA 90095-1547 USA}
\address{
2) Department of Physics
Technical University of Budapest,
Budapest, Hungary}
\address{
3) Department of Chemistry and Biochemistry
UCLA,
Los Angeles, CA 90095-1569 USA}

%\date{\today}
\maketitle

\begin{abstract}
Using one- and two-dimensional NMR spectroscopy applied to $^{13}$C spin-labeled \sasf\ and \spf, we demonstrate the existence of an intermediate charge-ordered phase in the TMTTF family of charge-transfer salts. At ambient temperature, the spectra are characteristic of nuclei in equivalent environments, or molecules. Below a continuous charge-ordering transition temperature T$_{co}$, the spectra are explained by assuming there are two inequivalent molecules with unequal electron densities. The absence of an associated magnetic anomaly indicates only the charge degrees of freedom are involved and the lack of evidence for a structural anomaly suggests that charge/lattice coupling is too weak to drive the transition.

PACS \#s 71.20.Rv, 71.30.+h, 71.45.Lr, 76.60.-k

\end{abstract}

\

]

\newpage

Recent evidence that electronic correlations can lead to inhomogenous charge and spin structures has become a dominant theme in analyzing the properties of doped transition-metal oxides such as the high-T$_c$ cuprates \cite{Tranquada1995} and the manganites \cite{Mori1998}. An important feature is that the details of the structures can fundamentally influence the low-temperature physics in ways that might otherwise seem inconceivable, possibly even the creation of a superconducting state by doping an antiferromagnetic insulator \cite{Kivelson1999}. 

\begin{figure}[htb]
\epsfxsize=.9\hsize
\epsffile{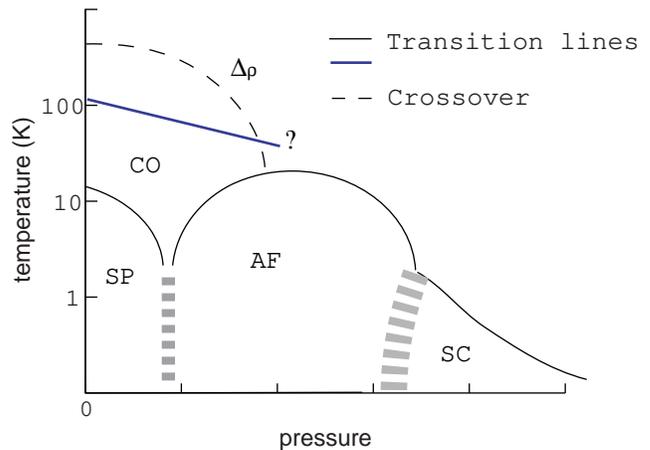}
\caption{Temperature vs. pressure phase diagram for the Bechgaard salts \seX, and the sulfur analogs \sX. The symbols are as follows: SP=Spin-Peierls, AF=antiferromagnetic, SC=superconductivity, $\Delta_{\rho}$=dimerization charge gap, and CO=charge-ordered. The solid lines are phase transitions, and the dashed line is a crossover. The hashed marks may be discontinuous transitions.}
\label{chargefg1}
\end{figure}

Observations~of charge-ordering in (DI-DCNQI)$_2$Ag \cite{Hiraki1998} and (BEDT-TTF)$_2$X \cite{Chiba2000,Takano2000} indicate inhomogeneities occur in some organic conductors as well. Prototypical among these is the family of isostructural TMTTF and TMTSF charge transfer salts \cite{Bourbonnais1999}. For nearly twenty years, the remarkable diversity of physical properties they exhibit have been summarized using a single temperature/pressure phase diagram (Fig. \ref{chargefg1}), where pressure is the parameter controlling the ratio of two competing energy scales. Note the existence of a superconducting phase next to an antiferromagnetic insulator \cite{Jerome1980}. Below, we describe observations which require the incorporation of a new transition line to the phase diagram (bold, blue line), below which the systems are charge-ordered (CO). The observation of an intermediate phase in this class of compounds can be explained by including a new energy scale \cite{Seo1997,Mazumdar1999}, and is particularly significant because the influence of the new scale can be examined across a range of ground states by pressure-tuning the system. And since there is evidence from independent transport measurements for CO fluctuations far above the CO transition temperature T$_{co}$, the interactions which drive the transition are relevant far into the normal phase and over a range of pressures \cite{Javadi1988,Nad2000}.

First we discuss Fig. \ref{chargefg1} while explicitly excluding the CO transition. The sequence of observed ground states (Spin-Peierls (SP), antiferromagnetic (AF), and superconducting (SC)) follows naturally from the combined effects of tunable dimensionality and on-site correlations \cite{Bourbonnais1999,Brazovskii1985,Bourbonnais1986}. Near to the SP ground state is a high-
\noindent symmetry phase with a charge gap ($\Delta_{\rho}$), whereas near to the SC ground state is a highly conducting normal state. Emery, \etal\ \cite{Emery1982} were the first to point out a simple mechanism by which this crossover from insulating to metallic behavior could occur without crossing a phase boundary, and their proposal led to the composite phase diagram. The compounds are formed by stacking the planar TMTTF molecules, and then lining up the stacks into layers that are separated by layers of counterions. If the molecular stacks are considered as weakly coupled chains with alternating intermolecular distances, then two-particle Umklapp processes produce a charge gap $\Delta_{\rho}$. Either an increase in transverse hopping or a decrease in the dimerization potential, as applied pressure would do, deconfines the charges and restores the conducting state. That is, application of pressure is equivalent to controlling the ratio of the dimerization gap to the transverse overlap integral ($\Delta_{\rho}$/t$_{\perp}$), which in turn determines the properties of the normal state. 

Here we demonstrate the existence of an intermediate, charge-ordered phase in \spf\ and \sasf, and propose that off-site Coulomb interactions are responsible. Strictly speaking, introducing a new energy scale modifies the physical properties exhibited by a particular compound, so the phase diagram of Fig. \ref{chargefg1} is better described as a slice of a diagram with at least one additional axis. Several previously-unexplained observations can be understood by recognizing the existence of the CO transition. 

Our conclusions are based on $^{13}$C NMR spectroscopy from samples of \spf\ and \sasf\ that were grown using standard electrolysis. Spin-labeled molecules were synthesized at UCLA \cite{Merlic1999} with the two 100\% $^{13}$C-enriched carbon sites forming the bridge of the TMTTF dimer molecule. All of the NMR measurements were made in an external field of B$_0$=9.00T, corresponding to an NMR frequency of 96.4MHz. 

\begin{figure}[htb]
\epsfxsize=.9\hsize
\epsffile{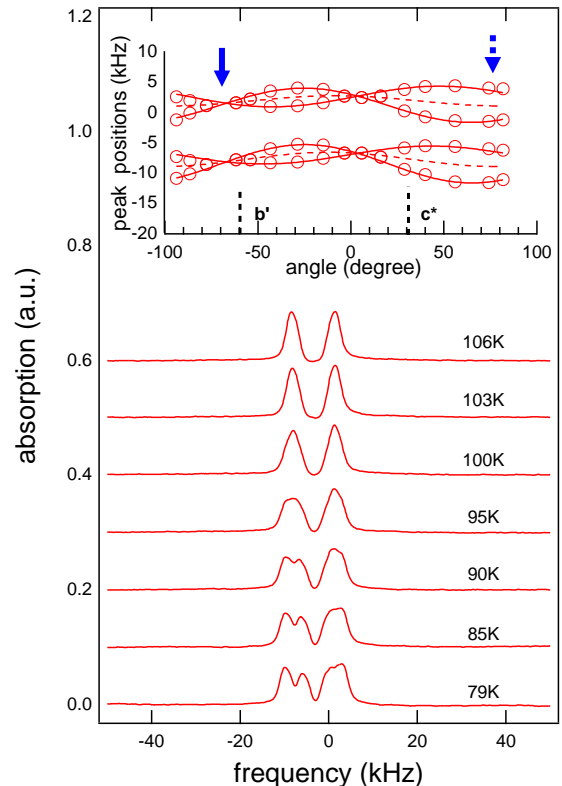}
\caption{$^{13}$C NMR spectra for (TMTTF)$_2$AsF$_6$ recorded at different temperatures. The inset shows the angular dependence of the spectrum at T=300K. A solid arrow denotes the angle at which the spectra in the main part of the figure were recorded. The dashed arrow refers to the angle associated to the data of Fig. \ref{chargefg4}.}
\label{chargefg2}
\end{figure}

In Fig. \ref{chargefg2}, seven 1D $^{13}$C NMR spectra for \sasf\ at representative temperatures are shown. At ambient temperature, each molecule is equivalent, but the two $^{13}$C nuclei in each molecule have inequivalent hyperfine coupling, giving rise to two spectral lines. The angular dependence of the spectral frequencies appears in the inset; the broken lines are the hyperfine shifts and the addition of a nuclear dipolar coupling gives the solid lines. The solid arrow is the angle at which the seven spectra were recorded.

Upon cooling, the NMR spectrum remains unchanged down to T=105K, below which each of the two peaks appear to split. From each molecule there is a signal from the nucleus with a stronger hyperfine coupling and a signal from the nucleus with a weaker hyperfine coupling. The doubling comes about because of two different molecular environments of roughly equal number, one with slightly greater electron density and one with a reduced electron density. Following the effects of the charge disproportionation to low temperature was difficult, because the SP fluctuations lead to line broadening and spectral overlap. However, we were able to use 2D J-resolved spectroscopic techniques to "unfold" unresolved signals from coupled nuclear spins. These measurements are discussed below.

The obvious choice for investigating the generality of the CO phenomenon is \spf, a system with physical properties originally used to identify Fig.\ref{chargefg1} as the appropriate phase diagram \cite{Bourbonnais1999,Brazovskii1985,Bourbonnais1986}, and recently found to be superconducting at a pressure of P$\approx$5.2 GPa \cite{Kobayashi1999}. In previous high-field $^{13}$C NMR spectroscopy on this compound, we had identified four inequivalent nuclei in the domain-walls of the incommensurate SP phase, rather than the expected two \cite{Brown1999}. The present results demonstrate that this is a consequence
\begin{figure}[htb]
\epsfxsize=1\hsize
\epsffile{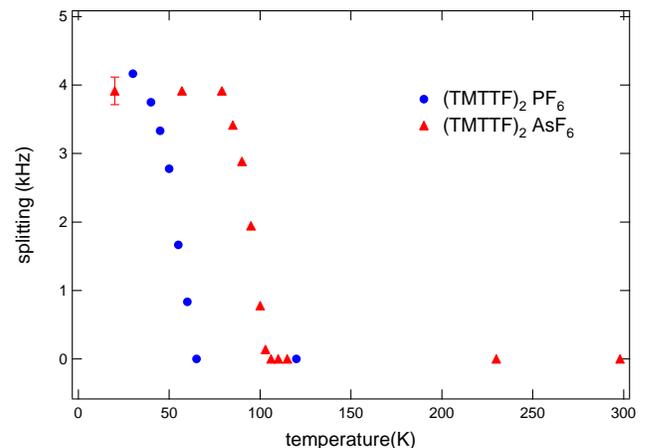}
\caption{Spectral splitting ($\sim$charge disproportionation order parameter) vs. temperature as obtained from 1D and 2D $^{13}$C NMR spectroscopy for two TMTTF-based salts.}
\label{chargefg3}
\end{figure}
\noindent of a charge-ordering occurring at a higher temperature. Even though the spectra were complicated by overlap, 2D J-resolved experiments led to unambiguous identification of a CO transition at approximately T=65K. The temperature dependence of the order parameter exhibited in Fig.\ref{chargefg3} shows that the transition is continuous to within the experimental resolution. Our measurements confirm the hypothesis put forward in recent reports of ac transport measurements, where a large and strongly frequency-dependent dielectric constant was attributed to the response of a charge-ordered phase \cite{Nad2000,Nad1999}.

An important puzzle of the TMTTF salts is solved by these experiments. It has been known for a long time that properties of certain TMTTF salts, for example (TMTTF)$_2$SbF$_6$, did not fit into the generally accepted model \cite{Coulon1982}. The temperature dependence of the resistivity $\rho$(T) for this material is metallic, that is, d$\rho$/dT$>$0 down to T=155K, where it appears that a continuous metal-insulator transition takes place \cite{Laversanne1984}. It was referred to as "structureless" because no signature was found in X-ray scattering studies. Also, the spin susceptibility is transparent to the structureless transition. Later, Coulon, {\it{et al.}} identified a feature in the thermopower of \sasf\ at T$\approx$100K, and through doping studies, they were able to establish that it was the same type of transition \cite{Coulon1985}. Taken together with our observations, the implication is that the structureless transition is a CO transition, it appears to be continuous, it is primarily the charge degrees of freedom which are involved, and it is ubiquitous to the TMTTF family. The charge-ordering is probably the reason why the activation energy obtained from transport measurements increases upon cooling, even though the dimerization decreases \cite{Pouget1996}. 

The mechanism for producing a CO phase is not unique, and both off-site Coulomb correlations \cite{Seo1997,Mazumdar1999}, and lattice coupling \cite{Mazumdar1999} are proposed as possible explanations. Although it is not proven, we believe that it is a consequence of long-range Coulomb interactions, because there was no previous identification of a structural anomaly, and we have seen no evidence for molecular reorientation in the 2D J-resolved experiments. Future scattering studies will be very helpful in this regard. 

There are implications for our understanding of the physical properties in the Bechgaard salts in particular and organic conductors in general. First, the details of the charge ordering in the insulators, and therefore the finite-ranged Coulomb interactions, can influence the 3D SP or AF orderings, including the pressure-tuned commensurate/incommensurate AF transition \cite{Klemme1995}. Also, CO fluctuations are evident from dielectric measurements to persist up to very high temperatures in a number of salts \cite{Javadi1988,Nad2000}, suggesting that dynamic density correlations are important to charge transport in the normal state. Of particular importance is the possibility that such correlations could be important for self-doping effects \cite{Vescoli1998}.

Lastly, we present some of the 2D spectra which helped us establish the temperature dependence of the charge disproportionation. The 2D J-resolved technique is useful in separating interactions along two frequency
\begin{figure}[htb]
\epsfxsize=1\hsize
\epsffile{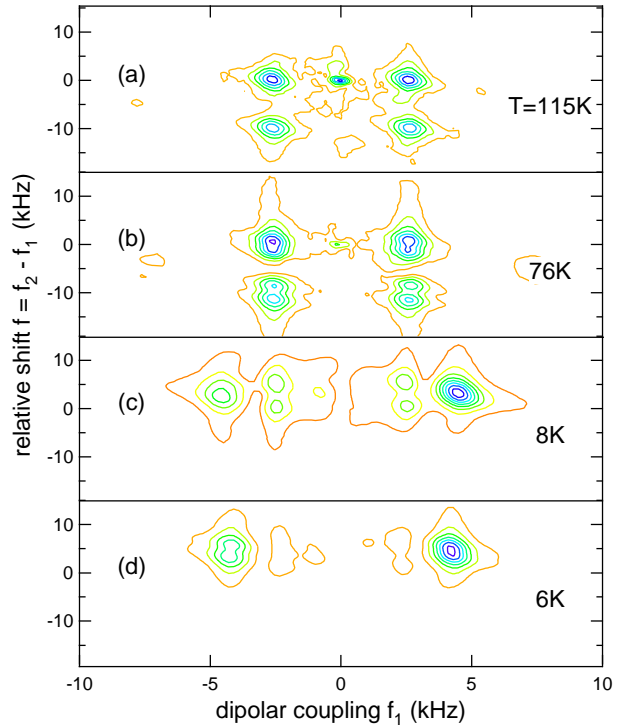}
\caption{2D J-resolved $^{13}$C NMR spectra from (TMTTF)$_2$AsF$_6$ at representative temperatures. a) T greater than the charge ordering transition temperature T$_{co}$. b) T$<$T$_{co}$. c) and d) T$<$T$_{SP}$. These spectra demonstrate the formation of a nuclear spin triplet state.}
\label{chargefg4}
\end{figure}
\noindent axes, hence providing more information than conventional 1D spectra, as well as sometimes resolving peaks that overlap \cite{Ernst1987}. Referring to Fig. \ref{chargefg4}, the spectra are shown with the nuclear dipolar interaction separated along the f$_1$ axis, and hyperfine interactions are separated along the f$_2$-f$_1$ axis. The orientation for these spectra are denoted by the the dashed arrow in the inset of Fig. \ref{chargefg1}.

The spectrum in Fig. \ref{chargefg4}a was obtained at T=115K; it arises from two spin I=1/2 nuclei seeing different hyperfine fields, coupled by the nuclear dipolar interaction. Repeating the experiment at T=76K, below the charge-ordering temperature T$_{co}$, leads to the charge-ordered spectrum shown in Fig. \ref{chargefg4}b, where each of the four peaks in Fig. \ref{chargefg4}a is doubled. No significant molecular reorientation occurs; that would change the dipolar coupling strength. By unfolding the spectra in this way, we were able to follow the charge ordering to lower temperatures where SP fluctuations lead to significant broadening.

The spectra in Fig. \ref{chargefg4}c) and d) are below the SP phase transition temperature T$_{SP}$(B=9T)=10.2K. These are displayed to demonstrate an interesting effect on the NMR spectrum associated with the quantum mechanics of the two coupled spins in each molecule. At these temperatures, the negative hyperfine shift has reduced significantly. In addition, an increase in the dipolar coupling energy by a factor of 3/2 is observed. Above the SP transition, the two carbon nuclei are "unlike" spins because of the different hyperfine shifts. The two coupled I=1/2 spins form four non-degenerate energy states. When the hyperfine shifts decrease for T$<$T$_{SP}$, the unlike spins become "like" spins and the appropriate basis states are the symmetric singlet and antisymmetric triplet. Only the triplet states are coupled by the nuclear dipolar interaction or the rf field, and further, the secular part of the internuclear dipolar interation increases by 3/2 \cite{Abragam1962}. In the data, the shift in spectral strength from the "unlike" to the "like" cases occurs smoothly as the difference in hyperfine interaction strength crosses below the characteristic dipolar interaction.

In~conclusion,~we have recorded 1D and 2D $^{13}$C NMR spectra in \sasf\ and \spf\ over the range 4-300K, from which we observe evidence of a charge-ordering transition. A direct connection between this phenomenon and the "structureless" transition is made, indicating that an intermediate, CO phase between the normal state and the ground state is ubiquitous in the TMTTF family of charge-transfer salts. The undetermined structure is essential information for identifying the mechanism, though it is probably caused by off-site Coulomb interactions. Dielectric experiments \cite{Javadi1988,Nad2000} demonstrate that CO fluctuations persist to very high temperatures, suggesting that a proper description of the high-symmetry phase will depend on including these effects. More generally, evidence for charge-ordering in a family of compounds that includes neighboring antiferromagnetic and superconducting phases is an important unifying observation in the field of highly correlated materials. 

ACKNOWLEDGEMENTS. We would like to recognize discussions with W. G. Clark, H. Fukuyama, H. Seo, S. Kivelson, S. Mazumdar, and P. Monceau. This work was supported in part by the National Science Foundation. 

%\bibliography{charge} \bibliographystyle{prstysb}

\end{document}